\documentclass[lettersize,journal]{IEEEtran}
\usepackage{amsmath,amsfonts}
\usepackage{algorithmic}
\usepackage{algorithm}
\usepackage{array}
\usepackage[caption=false,font=normalsize,labelfont=sf,textfont=sf]{subfig}
\usepackage{textcomp}
\usepackage{color}
\usepackage{stfloats}
\usepackage{url}
\usepackage{verbatim}
\usepackage{graphicx}
\usepackage{cite}
\usepackage{booktabs}
\begin{document}

\title{RefXVC: Cross-Lingual Voice Conversion with Enhanced Reference Leveraging}

\author{Mingyang Zhang,~\IEEEmembership{Member,~IEEE,}
        Yi Zhou,~\IEEEmembership{Member,~IEEE,} Yi Ren, Chen Zhang, Xiang Yin, 
        Haizhou Li,~\IEEEmembership{Fellow,~IEEE,}

\thanks{Mingyang Zhang, Haizhou Li are with the Shenzhen Research Institute of Big Data, School of Data Science, The Chinese University of Hong Kong, Shenzhen, China. (email: zhangmingyang@cuhk.edu.cn; haizhouli@cuhk.edu.cn)}
\thanks{Yi Zhou are with the Department of Electrical and Computer Engineering, National University of Singapore, Singapore (e-mail: yi.zhou@u.nus.edu)}
\thanks{Yi Ren, Chen Zhang, Xiang Yin are with Speech \& Audio team, ByteDance AI Lab (email: ren.yi@bytedance.com, zc99@zju.edu.cn, yinxiang.stephen@bytedance.com)}
}

\markboth{Journal of \LaTeX\ Class Files,~Vol.~14, No.~8, August~2021}%
{Shell \MakeLowercase{\textit{et al.}}: A Sample Article Using IEEEtran.cls for IEEE Journals}


\maketitle

\begin{abstract}
This paper proposes RefXVC, a method for cross-lingual voice conversion (XVC) that leverages reference information to improve conversion performance. Previous XVC works generally take an average speaker embedding to condition the speaker identity, which does not account for the changing timbre of speech that occurs with different pronunciations. To address this, our method uses both global and local speaker embeddings to capture the timbre changes during speech conversion. Additionally, we observed a connection between timbre and pronunciation in different languages and utilized this by incorporating a timbre encoder and a pronunciation matching network into our model. Furthermore, we found that the variation in tones is not adequately reflected in a sentence, and therefore, we used multiple references to better capture the range of a speaker's voice. The proposed method outperformed existing systems in terms of both speech quality and speaker similarity, highlighting the effectiveness of leveraging reference information in cross-lingual voice conversion. The converted speech samples can be found on the website: \url{http://refxvc.dn3point.com} 
\end{abstract}


\section{Introduction}
Among speech synthesis tasks, cross-lingual voice conversion (XVC) is an interesting research topic that allows for the conversion of the speaker's voice from one language to another while maintaining the speaker's identity \cite{abe1990cross,charlier2009cross,sundermann2006textcross,zhou2019cross}. For instance, XVC enables the actor in a Hollywood English movie to speak perfect Spanish/Mandarin/Hindi/etc. It is challenging, but the enabling technology for various real-life applications, e.g., foreign language education \cite{abe1990cross}, speech-to-speech translation \cite{erro2010voice}, foreign movie dubbing, and so on \cite{zhou2019cross}. In the XVC task, converted speech is expected to be sound as being pronounced by a native speaker.

If there are a number of high-quality speech data from the target speaker, one can easily build an acoustic model with the data for conversion. However, it is usually unrealistic to obtain such a quantity of speech data for each of the target speakers. Zero-shot XVC, therefore, draws the researcher's attention, where only a few speech samples are required to enroll for generating desired voice. Thanks to the power of deep learning, zero-shot XVC has achieved great success. It generally takes speech data from multiple speakers and generates desired voice by conditioning on a speaker identity representation. A speaker can be simply represented by a fixed vector produced by pre-trained neural speaker recognition models \cite{variani2014deep} or a disentangled speaker embedding obtained from an encoder module \cite{wang2023neural}. Such speaker representations usually carry the averaged features per speaker \cite{variani2014deep} or per utterance\cite{wang2023neural}. 

The main motivation behind XVC is to enable seamless communication between speakers of different languages while preserving the naturalness and identity of the target speaker. However, XVC is a challenging task because it requires the conversion of not only the phonetic and linguistic features of the source speech but also the speaker identity and prosodic features \cite{10109809}. One of the major challenges in XVC is to deal with the timbre changes that occur when a speaker produces different pronunciations across languages. Previous works in XVC have generally used an average speaker embedding to represent the speaker's voice, which does not account for these timbre changes. As a result, the converted speech may sound unnatural or contain artifacts \cite{9606610}.

In addition to the timbre problem, a connection between timbre and pronunciation in different languages can be observed. Therefore, it also requires the conversion model to leverage the content information of the references to improve the converted speech quality\cite{Pronunciation}. However, it is found that the variation in tones is not adequately reflected in a sentence, and therefore, multiple references are necessary to better capture the range of a speaker's voice.

Furthermore, The task of XVC involves two language systems. The differences in pronunciation, intonation, and other linguistic features create further obstacles to accurately describing the vocal tract of a speaker. Consequently, dissimilar traces towards the actual speaker still exist in the converted speech \cite{8911462}.

%
%


To address the challenges, we present the RefXVC system, which seeks to leverage speaker information from the reference to the maximal extent in order to improve XVC performance. Our proposed XVC network utilizes the autoencoder architecture to map input self-supervised learning (SSL) representations to acoustic features, which are conditioned on fine-grained speaker embeddings extracted using several techniques. Our approach is centered around the following key aspects:

1. We introduce a timbre encoder that extracts both global and local speaker embeddings from the source speech. They are combined to capture the time-varying speaker characteristics in a given sentence. The global speaker embedding characterizes the overall characteristics of the speaker's voice, while the local speaker embeddings represent the fine-grained variations in timbre that occur with different pronunciations. By using a timbre encoder to extract both types of speaker information, we can ensure that the synthesized speech has the correct timbre and tone of the target speaker, which can improve the naturalness and authenticity of the generated speech.

2. The second aspect of our approach is the design of a pronunciation matching network to utilize content-related speaker information. The pronunciation matching network is trained to align SSL features of the source speech with those of the reference speech. By using content-related speaker information, such as the pronunciation of specific words and phrases, we can ensure that the converted speech has the correct pronunciation of the source sentence, which can improve the intelligibility and accuracy of the converted speech.

3. We further employ the use of multi-reference encoding to enrich the content information. In many cases, a single reference speech may not contain enough information to cover the nuances and variations of the source speech. Using multiple reference speech samples, we can enrich the content information and ensure that the converted speech has the correct intonation of the source content. This can improve the naturalness and expressiveness of the converted speech, which can be particularly important in applications of XVC.

Overall, our method tackles the challenges of XVC by utilizing a timbre encoder to extract both global and local speaker embeddings, a pronunciation matching network to utilize content-related speaker information, a multi-reference encoding technique to enrich the content information, and a normalized pitch as input to better preserve the native prosody. By combining these techniques, we can improve the quality of the converted speech and make it sound more natural and similar to the target speaker.

In the experiment, we convert the voice between English and Spanish speakers. We verify that the proposed system synthesizes natural speech with high speaker similarity by prompting in the zero-shot XVC task.

\section{Related Works}
In this section, we revisit several voice conversion frameworks and self-supervised learning representations for zero-shot XVC. We also study the related works in speaker code representation.

\subsection{Cross-Lingual Voice Conversion}
In cross-lingual voice conversion (XVC) tasks, source refers to an utterance from a native speaker in one language, while target is defined as the utterance from another speaker who speaks a different language. Ideally, converted speech carries the source's speech content while presenting the target's timbre. 
Popular XVC frameworks generally adopt the encoder-decoder architecture to disentangle the speaker-dependent component (speaker identity) from the speaker-independent component (speech content). In this way, one can convert the voice from one another by just changing the speaker-dependent component while keeping the speaker-independent component. 

Variational autoencoder (VAE)~\cite{kingma2019introduction,hsu2017voice,tjandra2019vqvae, app132111988} is one such implementation, which learns a latent space for speaker-independent representation. 
Similarly, generative adversarial networks (GAN) \cite{goodfellow2020generative} disentangle the speech attributes with an extra adversarial loss to guarantee a distribution match between the generated and true data \cite{kaneko2019cyclegan,sisman2019study}. AutoVC is designed with a carefully designed bottleneck to constrain the information flow and is equipped with a speaker verification module to learn the speaker embedding \cite{qian2019autovc}. Whisper\cite{pmlr-v202-radford23a} is also employed as a content feature extractor for XVC in \cite{10389651}. The authors introduced the speaker consistency loss to enhance the speaker information contained within the extracted speaker embedding. 
However, these methods also do not explicitly leverage linguistic content as supervision, thus sometimes producing unclear or distorted samples.

Alternatively, pretraining techniques greatly enhance the conversion performance by providing a linguistic representation learned by a well-trained model. Automatic speech recognition (ASR) is a perfect option, which can provide linguistic representations by either Phonetic PosteriorGram (PPG)~\cite{hazen2009query,sun2016phonetic,zhou2019cross}, 
or directly the discrete output phonemes. Yet, this requires extensive labeled data for training the ASR models. Additionally, recognition errors highly affect the generated speech intelligibility.

\subsection{SSL Representation}

Self-supervised learning (SSL) representation has gained significant attention in recent years due to its ability to efficiently learn meaningful features from large and unstructured datasets \cite{misra2020self}, which is especially useful in domains such as computer vision \cite{tung2017self}, natural language processing \cite{lan2019albert}, and speech recognition \cite{ravanelli2020multi}. 
%
%
Several well-known SSL representations, e.g., Autoregressive Predictive Coding (APC) \cite{chung2020generative}, Contrastive Predictive Coding (CPC) \cite{oord2018representation}, and wav2vec \cite{baevski2020wav2vec} have been studied for voice conversion \cite{huang2022s3prl,chung2020generative,riviere2020unsupervised}.

\subsubsection{HuBERT Token}
The HuBERT token is a special type of token that is used in the HuBERT model \cite{hsu2021hubert}. It is a self-supervised pretraining model for speech processing similar to the popular BERT model. The model learns to predict masked portions of the audio signal. The HuBERT token represents the start and end of each audio clip used in the model. This is important because speech processing tasks often require analyzing long audio sequences, and the HuBERT token helps the model to segment the audio into smaller, more manageable segments. By doing so, it encodes meaningful representations of the audio signal in a way that can be useful for a variety of speech processing tasks, such as speech recognition \cite{hsu2021hubertasr}, speaker identification \cite{wang2021fine}, and speech-to-speech translation \cite{lee2021textless}.

The HuBERT token is advantageous over other token-based approaches because it enables the model to process speech directly without the need for any additional preprocessing. This means that the model can learn speech features more efficiently and accurately. Additionally, the HuBERT model is designed to handle long sequences of speech, which is critical for many speech generation tasks. It has been investigated in XVC and obtained impressive performance \cite{ren2023bag}; hence, the HuBERT representation is an ideal option to set our starting point. 

\subsection{Neural Speaker Encoding}
The use of HuBERT tokens makes the content representation a fixed embedding. The choice of speaker representation is critical to obtain the desired voice. There are two popular approaches to obtain a speaker embedding: 1) extracting the hidden representations from pre-trained a speaker recognition systems such as \textit{d-vector} \cite{variani2014deep}, \textit{x-vector} \cite{snyder2018x}, and ECAPA-TDNN \cite{desplanques2020ecapa}; 2) jointly training a speaker encoder through multitask learning and extracting bottleneck features as a speaker embedding. The latter method typically involves disentangling the speaker information with an adversarial loss on the speaker classification task \cite{chen2021again,wu2020one,chou2019one}.

Pretraining a speaker recognition system offers the advantage of using large-scale speaker databases, enabling the learned speaker representation to exhibit high speaker similarity in several multi-speaker speech generation frameworks~\cite{jia2018transfer,qian2019autovc,zhang2021transfer,zhou2021language}. On the other hand, joint training provides a more flexible optimization process dedicated to the speech synthesis task, providing further insights to characterize the speaker details \cite{ding2020improving,huang2020unsupervised}.

Resemblyzer is a popular choice in recent speech synthesis studies that allows for deriving a high-level representation of a voice through a deep learning model. Given an audio file of speech, Resemblyzer\footnote{\url{github.com/resemble-ai/Resemblyzer}} creates a summary vector of 256-dimensional embedding that captures the voice's characteristics. This is taken as a suitable reference for this work.

\subsection{Personalized Speech/Singing Voice Synthesis}
Personalized text-to-speech (TTS) and singing voice synthesis (SVS) techniques generate speech and singing, respectively. They share the same objective with voice conversion of generating realistic and natural-sounding voices in a specific speaker's voice, while the input of TTS and SVS are text and musical information, respectively, instead of speech. Current TTS and SVS frameworks mainly adopt the encoder-decoder architecture, where the encoder projects input into a latent embedding, and the decoder predicts acoustic features conditioned on a speaker representation \cite{ping2017deep,jia2018transfer,blaauw2017neural,cho2021survey}. 

It can be noted that both TTS and SVS decoders work in the same way as the decoder in the VC system decoder, which relies on speaker embeddings to vary the speaker identity. Hence, their efforts in the speaker encoding adoption serve as a valuable source of inspiration for our research in XVC. For example, the TTS work described in \cite{choi2022nansy} attempts to encode timbre information using a content-dependent time-varying speaker embedding. The model successfully captures timbre information of unseen target speakers during training. On the other hand, several SVS studies \cite{zhang2020durian,lee2020disentangling} have established a correlation between speech and singing and demonstrated the benefits the learning of a unified speaker representation. Moreover, certain multilingual TTS works \cite{nekvinda2020one} have incorporated a secondary fine-tuning step to optimize a speaker identity-preserving loss, enabling the model to output a consistent voice regardless of language. These methods have a common goal of encouraging sharing of model capacity in speaker representation learning across linguistic and prosodic variations \cite{black2004multilingual,azizah2020hierarchical}. 


These findings support our assumption that it is crucial to discover and establish the relationships in one's vocal tract while pronouncing various content in different. The resultant speaker representation should be robust and comprehensive against languages, which is ideal for XVC. 

\begin{figure*}[!ht]
    \centering
    \captionsetup[subfloat]{labelfont=footnotesize,textfont=scriptsize}
    \subfloat[RefXVC]{\includegraphics[width=.45\textwidth]{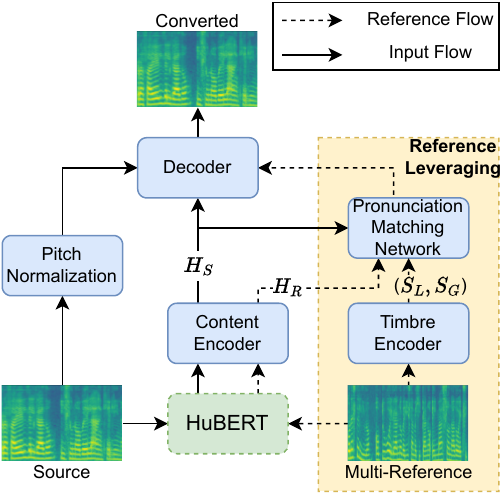}
    \label{fig:all}}
    \hspace{1.5cm}
    \subfloat[Pronunciation Matching Network]{\includegraphics[width=.25\textwidth]{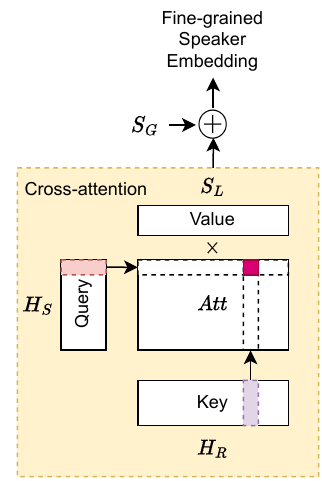}
    \label{fig:pmn}}
    \caption{(a) The overall architecture of the proposed RefXVC. (b) The details of the pronunciation matching network. $H_S$ denotes the hidden representation of source HuBERT, $H_R$ denotes the hidden representation of reference HuBERT, $S_L$ denotes the local speaker embedding and $S_G$ denotes the global speaker embedding.}
    \label{fig:system}
\end{figure*}

\section{Proposed R\MakeLowercase{ef}XVC}
This section introduces the proposed RefXVC framework, including introducing the flow-based XVC framework, multi-lingual HuBERT token, timbre encoder, pronunciation matching network, multi-reference encoding, and pitch normalization.

\subsection{Flow-Based XVC Framework}
The proposed RefXVC model is based on an autoencoder-like architecture that employs several information extraction modules to generate disentangled representations of Mel-spectrograms, as illustrated in Figure \ref{fig:all}. These modules consist of the following components: a content encoder that encodes input SSL representations into latent content information, a timbre encoder that extracts the target speaker's timbre information from the reference speech, a pitch normalization module that normalizes the pitch on a per-sentence basis, a pronunciation matching network that represents a fine-grained time-varying timbral representation, and a speech decoder that takes all representations as input to generate speech with source content and target voice.

The content encoder and speech decoder follow the work presented in \cite{ren2023bag}. The content encoder is a stack of feed-forward Transformer layers with relative position encoding. And the speech decoder reconstructs the Mel-spectrogram using a mean absolute error (MAE) and a multi-length adversarial loss. The next sections will provide more details about each of the other sub-modules.

\subsection{Multilingual HuBERT Token}
SSL pretraining has demonstrated its strength in learning high-level feature representations for various downstream tasks, where speech processing is one successful instance. In this section, we introduce the use of Multilingual HuBERT Token, a self-supervised learning model, to extract the feature representation from the reference speech. This multilingual model is capable of extracting phonetic features from input speech, irrespective of the language being spoken. This SSL pretraining allows us to leverage the massive amount of unlabeled speech data across different languages to learn a more robust and comprehensive feature representation.

We first encode the input speech using HuBERT to obtain its SSL representations, which provide a high-level representation of the input speech by capturing its phonetic and acoustic features. We then use this SSL representation as input to our XVC network. The multilingual HuBERT projects speech in different languages onto a common feature space representing the linguistic information and serves as a bridge between languages, thus enabling XVC.

Additionally, this SSL representation is also taken as input to our pronunciation matching network. It allows the network to focus on the phonetic features of the input speech and match them with corresponding features in the reference speech. This ensures accurate pronunciation and reduces foreign accents. The use of HuBERT is essential to handle XVC with high efficiency and accuracy.

\subsection{Timbre Encoder}
The timbre of a speaker's voice can vary depending on attributes such as pronunciation and intonation. Previous works commonly used an average speaker embedding to represent the speaker information, which may not capture the time-varying characteristics of the speaker's voice details. To tackle this problem, we suggest utilizing a timbre encoder to extract global and local speaker embeddings. This approach aims to combine both representations to accurately characterize dynamic speaker information that varies over time, which is expected to enhance the preservation of speaker identity and the quality of the converted speech in XVC tasks.
These embeddings can be used to benefit the following modules of the system, such as the multi-reference encoding and pronunciation matching network, by providing additional speaker-related information.

The timbre encoder is a critical component in our proposed XVC system. It is responsible for extracting the speaker embeddings, which are used to characterize the target speaker. We refer to this module as the "timbre" encoder because it captures the timbral characteristics of the speaker's voice. To extract both utterance-level and frame-level speaker embeddings, we use a three-layer {bidirectional} LSTM neural network. The speaker embedding is obtained by passing the entire reference speech through the LSTM network. The resulting output from the last LSTM layer is used as the utterance-level or global speaker embedding $S_G$, which characterizes the speaker's overall voice characteristics. The output of the last LSTM layer at each frame is used as the frame-level or local speaker embedding $S_L$, which characterizes the speaker's voice characteristics at that specific moment in time. By extracting both utterance-level and frame-level speaker embeddings, our timbre encoder captures both the overall characteristics of the speaker's voice as well as the finer nuances of their speech. These speaker embeddings are then used as inputs to the following modules, which generate the converted speech that matches the content of the input speech while maintaining the speaker's characteristics.

\subsection{Pronunciation Matching Network}
In this section, we introduce the proposed Pronunciation Matching Network (PMN), which aims to leverage the content information from the reference speech during training. As discussed earlier, the timbre of a voice in voice conversion tasks is not constant and changes with different pronunciations. Previous works mainly relied on averaging speaker embedding, which is fixed at the utterance level or even for a speaker. Consequently, this embedding could fail to capture the dynamic variation in timbre. Additionally, in XVC, we believe a correlation exists between timbre and pronunciation in different languages, which has not been fully established in previous works. In this work, we would like to explore and propose PMN to reveal the unobtrusive relation between SSL representations of the source and reference speech by considering the pronunciation similarity of different languages. In this way, the network is empowered with an enhanced leveraging capability of characterizing the dynamic content information from the reference speech. The resultant conversion performance is expected to be robust over both speech quality and speaker identity preservation. 

Cross-attention \cite{NIPS2017_3f5ee243} is a powerful technique used in natural language processing and deep learning. It allows neural networks to attend to multiple parts of the input sequence simultaneously and weigh the importance of each part differently. This is achieved by computing attention scores between different positions of the input sequence and then combining the outputs of these computations to create a single output vector. Cross-attention has been used in a variety of tasks, including machine translation, image captioning, and question-answering. It has achieved significant improvements over traditional attention mechanisms. The ability to model complex relationships between different parts of the input sequence makes it a key component in many state-of-the-art models in natural language processing.

In this work, we introduce a novel PMN module to address the issue of speaker variability in XVC, as illustrated in Figure \ref{fig:pmn}. The network seeks to represent a fine-grained time-varying timbral representation dedicated to XVC tasks in a content-aware manner. This network enables the model to learn the optimal alignment between the input and reference speech, leading to improved speaker similarity in the generated speech. We incorporate a content-aware cross-attention module into the XVC network to achieve this. The cross-attention module takes the hidden representation of the source HuBERT tokens $H_S$ as queries, the hidden representation of the reference HuBERT tokens $H_R$ as keys, and the frame-level speaker embedding $S_L$ extracted from the reference utterances as values. The cross-attention mechanism allows the model to learn the alignment between the source and reference speech and to extract the fine-grained speaker embedding that is applied to the utterance-level speaker embedding $S_G$. The PMN is designed to learn the phonetic features of the source and reference speech and to identify the similar pronunciation units between the two languages. By using the frame-level speaker embedding extracted from the reference speech, PMN can capture the subtle variations in speaker characteristics that are critical for speaker similarity in XVC.

\subsection{Multi-Reference Encoding}
In many real-world scenarios, obtaining a reference speech that contains similar pronunciation to the source input is not trivial. Even if a reference speech is available, it may not be enough to capture all the acoustic variations that exist in the source speech. Moreover, it is easy to obtain several sentences or utterances spoken by the source speaker as reference speech during inference. To address these challenges, we introduce a multi-reference encoding technique that can effectively leverage multiple references to improve cross-lingual voice conversion performance.

Multi-reference encoding is a technique used in natural language processing and information retrieval to improve the accuracy of language models by incorporating multiple reference sources \cite{zheng2018multi}. Traditionally, language models use a single reference document or text corpus to learn patterns and generate predictions. However, this approach can be limited as it may not capture the full diversity and complexity of the language. Multi-reference encoding aims to overcome this limitation by encoding multiple reference sources simultaneously, allowing the model to learn from a wider range of examples and improve its ability to generalize to new inputs. This technique has been applied to a range of speech processing areas, including style transfer \cite{bian2019multi, Whitehill2020}, singing voice synthesis \cite{wang2022mr}, and speech recognition \cite{7404847}, and has led to significant improvements in performance. Multi-reference encoding is a promising area of research in natural language processing and is expected to continue to play an important role in advancing the field.

In this study, we tackle the challenge of employing a solitary reference speech in XVC by proposing a multi-reference scheme for network training. Rather than relying on a lone reference speech, we leverage multiple reference speech samples to enhance the content information and account for nuanced pronunciation variations of the target speaker. Our approach enables the XVC network to optimize the alignment between languages and elevate the overall quality of the converted speech.

To ensure consistent speaker identity across the entire reference speech, we use the timbre encoder to extract the global speaker embedding $S_G$ for each reference utterance. We introduce a speaker similarity loss that sums the cosine embedding loss for any two $S_G$:
\begin{equation}
    L_{ss} = \sum_{i,j\in N}cel(S_{G_i}, S_{G_j})
\end{equation}
where $N$ is the number of references and $cel()$ represents the cosine embedding loss.
This loss encourages the embeddings for the same speaker to be close together in the embedding space, making the speaker identity more consistent and improving the overall quality of the converted speech. We chose to use three utterances of the reference speech during training, setting $N = 3$ in our work. This is due to the GPU memory limitation, as using a larger $N$ would require more memory to store and process the additional reference utterances. By setting $N = 3$, we balance the trade-off between using enough reference information to improve the conversion quality and keeping the memory consumption within feasible limits. Furthermore, using more than 3 reference utterances may not lead to further improvement in conversion quality, as the benefit of additional reference information could saturate after a certain point.

During training, we also investigated the efficiency of our approach when multiple references are used without involving the source input. When we use references during training, we want to provide the model with additional information about the target speaker's voice without biasing it toward the source speaker's voice. If we were to include the source input as one of the reference utterances, the model might simply learn to copy the source speaker's voice rather than learn to generate a new voice that is similar to the target speaker's voice. By using multiple reference utterances that do not include the source input, we encourage the model to learn a more general mapping from the input speech to the target speaker's voice. This can help to reduce the risk of over-fitting to a particular reference utterance and can also help to capture more variation in the target speaker's voice.

\subsection{Pitch Normalization}

\begin{figure}
    \centering
    \includegraphics[width=.45\textwidth]{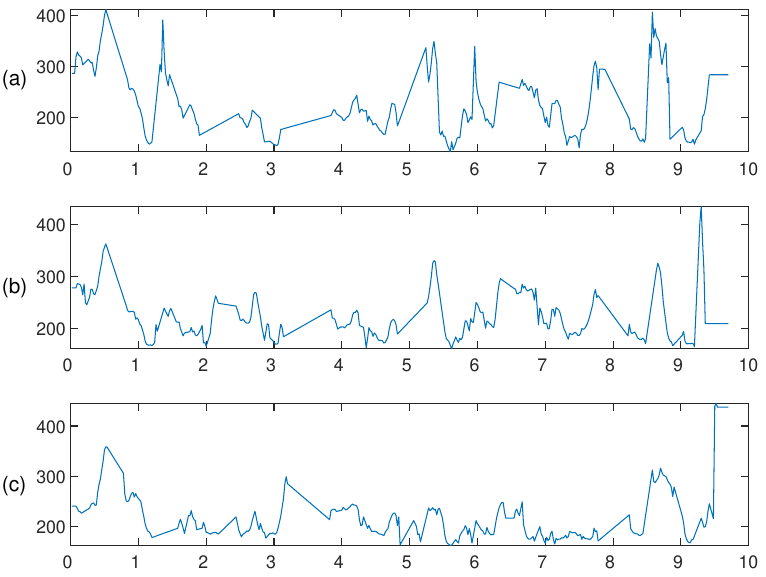}
    \caption{Illustration of the impact of normalized F0 on the system. (a) F0 of the source speech; (b) F0 of the converted speech utilizing normalized pitch; (c) F0 of the converted speech without utilizing normalized F0. The vertical axis stands for the amplitude of F0 in Hz, and the horizon axis stands for the time in seconds.}
    \label{fig:f0_com}
\end{figure}

In XVC, one of the main challenges is to ensure that the converted speech retains the prosodic characteristics of the source language. Prosody, which includes pitch, rhythm, and intonation, plays an important role in conveying emotions and meaning in speech. However, the pitch distribution of speakers varies widely depending on factors such as age, gender, and speaking style, and this can cause the converted speech to sound foreign or unnatural.

To address this issue, we propose to introduce normalized pitch as an additional input to the XVC system \cite{4317579, 9054734}. This enables explicit control of prosody in the converted speech, making it more similar to the input and reducing foreign accents. By normalizing the pitch on a per-sentence basis, we can ensure that the output speech has a similar prosody to the input, which is especially important in XVC tasks where a foreign accent can be a major issue. Overall, this approach allows for greater flexibility and control in XVC, ensuring that the output speech is not only recognizable but also natural-sounding and fluent.

We then incorporate the normalized pitch values as an additional input to the decoder module of the XVC network. The decoder takes the content information and fine-grained speaker embedding extracted from the timbre encoder, as well as the normalized pitch values as input, to generate the converted speech.

The F0 contour of the source speech and converted speech is compared in Figure \ref{fig:f0_com}, with and without the use of normalized pitch as an additional input. It can be observed that utilizing the normalized pitch as an additional input leads to the converted speech better following the prosody of the source speech, resulting in improved native sound and reduced foreign accent.

\section{Experiment}
\subsection{Database}
Our proposed system was evaluated on the Spanish-English XVC dataset through a series of experiments presented in this paper. For the English dataset, we utilized the train-clean-360 subset of the LibriTTS corpus, consisting of 191.29 hours of speech data from 904 speakers, including 430 female and 474 male speakers. Similarly, we used the Spanish subset of the Multilingual LibriSpeech (MLS) dataset for the Spanish dataset, which contains 917.68 hours of speech data from a total of 86 speakers, including 50 female and 36 male speakers. The data split was 90\% for training and 10\% for validation. For evaluation purposes, we employed the VCTK dataset \cite{veaux2017cstr} for English and the M-AILABS Speech Dataset \cite{Solak_2021} for Spanish. Allowing us to assess the performance of our method on unseen speakers and verify its generalization capability.  We extracted the 80-dimensional Mel-spectrum features and HuBERT tokens from all speech data by downsampling it to $16 kHz$, with a 20ms frameshift and a $64$ ms frame length.

\subsection{Model Architecture}
The content encoder and decoder architectures in this work are based on the VC system presented in \cite{ren2023bag}. The content encoder comprises a feed-forward Transformer \cite{NIPS2017_3f5ee243} with relative position encoding and a hidden size of 192. The decoder is composed of a posterior encoder, a speech decoder, and a multi-length discriminator. The posterior encoder consists of a 1D-convolution layer with stride 4, followed by ReLU activation and layer normalization, and a non-causal WaveNet layer. The number of encoder layers, WaveNet channel size, and kernel size are 8, 192, and 5, respectively. The speech decoder consists of a non-causal WaveNet layer and a 1D transposed convolution layer with stride 4, also followed by ReLU and layer normalization. The number of speech decoder layers, WaveNet channel size, and kernel sizes are set to 4, 192, and 5, respectively. The multi-length discriminator is an ensemble of three CNN-based discriminators that evaluate the Mel-spectrogram based on random windows with lengths of 32, 64, and 128 frames. Each CNN-based discriminator consists of $N + 1$ layers of 2D convolutions, each followed by a Leaky ReLU activation and a dropout layer. The latter $N$ convolutional layers are additionally followed by an instance normalization layer. After the convolutional layers, a linear layer projects the hidden states of the Mel-spectrogram slice to a scalar that represents the prediction of whether the input Mel-spectrogram is true or fake. In our experiments, we set $N = 2$ and the channel size of these discriminators to 32.

The timbre encoder is a 3-layer bidirectional LSTM network that takes Mel-spectrogram as input and generates a 256-dim utterance-level and frame-level speaker embedding. The pronunciation matching network is a cross-attention module composed of feed-forward Transformer layers. We also utilize a neural vocoder, HiFi-GAN, to convert the Mel-spectrogram to the waveform \cite{NEURIPS2020_c5d73680}.

We trained our model on a single NVIDIA GeForce RTX 3090 GPU, with a batch size of 16 by using the Adam optimizer with $\beta_{1}=0.9, \beta_{2}=0.98$, and the initial learning rate was 0.002 with the Noam decay scheme.

\begin{figure}
    \centering
    \captionsetup[subfloat]{labelfont=footnotesize,textfont=footnotesize}
    \subfloat[Speaker embedding from Resemblyzer]{\includegraphics[width= .38\textwidth]{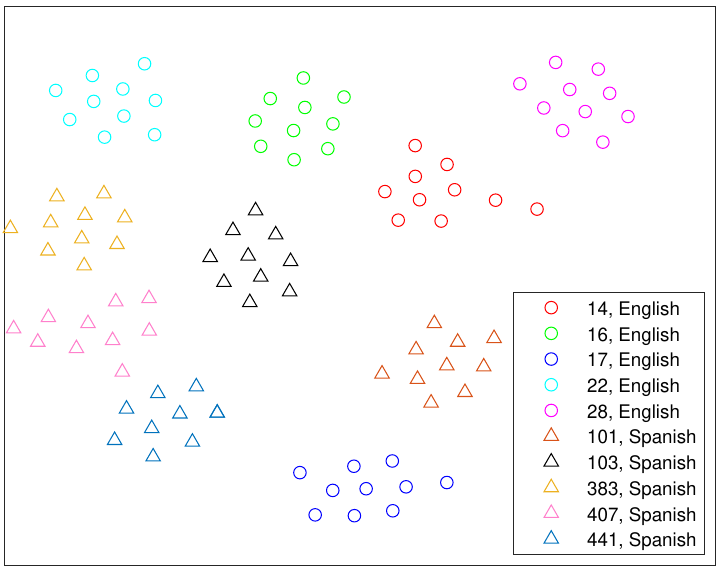}}\\
    \subfloat[Speaker embedding from the timbre encoder]{\includegraphics[width=.38\textwidth]{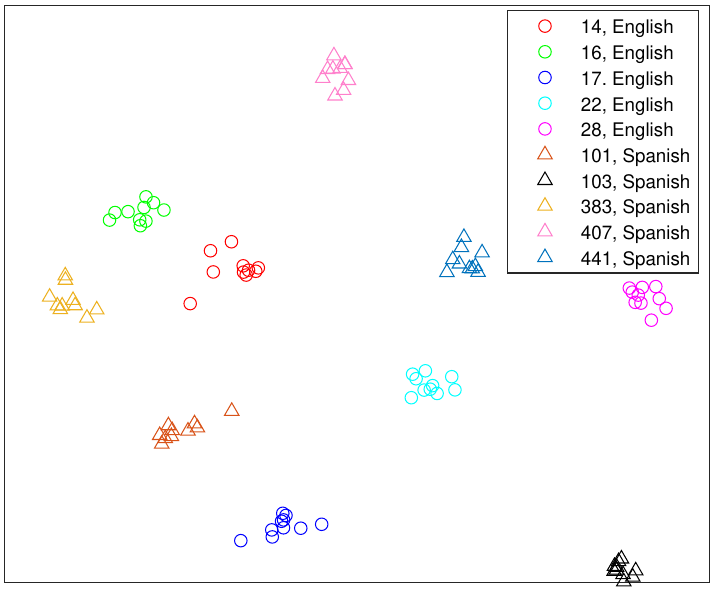}}
    \caption{Speaker embedding visualization using t-SNE. (a) speaker embedding obtained from Resemblyzer; (b) speaker embedding from the timbre encoder. Colors and shapes represent the speaker and language, respectively. The numbers are the identification codes of speakers in the database.}
    \label{fig:tsne}
    \vspace{-3mm}
\end{figure}

\subsection{Evaluations}

\subsubsection{Evaluation on Speaker Embedding}
In this section, we aim to evaluate the performance of the speaker embedding generated by the timbre encoder of our proposed system. We use t-SNE \cite{van2008visualizing}, a dimensionality reduction technique, to visualize the speaker embedding spaces, providing a qualitative and intuitive understanding of how our system operates.

To compare our speaker embedding with the state-of-the-art deep learning-based voice encoder, we used Resemblyzer \cite{g2e}, which has been widely adopted in voice cloning projects. As shown in Figure \ref{fig:tsne}, the speaker embedding space generated by Resemblyzer has a clear boundary for different languages. This means that speakers of different languages are represented differently in the speaker embedding space, which can make it challenging for XVC systems to preserve the speaker identity when converting between languages. In contrast, the speaker embeddings generated by our proposed system are language-agnostic, meaning that speakers from different languages are represented more similarly in the speaker embedding space. This makes our system more suitable for XVC across different languages, which is an important consideration for real-world applications.

Besides its language-agnostic property, we also analyzed the speaker embedding space produced by our proposed system and Resemblyzer in terms of intra-speaker and inter-speaker distances. Intra-speaker distance refers to the distance between embeddings of different utterances from the same speaker, whereas inter-speaker distance measures the distance between embeddings of different speakers. The analysis revealed that the speaker embedding space generated by our system has a smaller intra-speaker distance and a larger inter-speaker distance than Resemblyzer. This indicates that our system generates more compact speaker embeddings for the same speaker, enabling better differentiation between different speakers. In contrast, Resemblyzer tends to have larger intra-speaker distances, which may lead to less consistency in representing the same speaker.

The results of our evaluation demonstrate that our proposed system's timbre encoder generates a more suitable and effective speaker embedding space for XVC than Resemblyzer. The language-agnostic property of our speaker embedding space and its ability to capture the unique speaker characteristics make it more appropriate for XVC tasks, where speaker similarity across languages is crucial.
\subsubsection{Algorithm Comparison}
We conducted experiments to evaluate the performance of our proposed system in two source-to-target speaker conversion settings: English-to-Spanish and Spanish-to-English. For each setting, we considered four gender-to-gender combinations: male-to-male (m2m), male-to-female (m2f), female-to-male (f2m), and female-to-female (f2f). We selected two females and two males from each language, resulting in 16 ($=2\times2\times4$) conversion pairs for each setting.

We evaluate the conversion results of the following systems:
\begin{itemize}
    \item \textbf{Baseline}: XVC network with only content encoder, timbre encoder, and decoder. Only global speaker embedding is used, and PMN is excluded.
    \item \textbf{Single-RefXVC}: Our proposed RefXVC system with a single reference.
    \item \textbf{RefXVC (source-included)}: Our proposed RefXVC system with multi-reference technique while the source is included in the references during training.
    \item \textbf{RefXVC (source-excluded)}: Our proposed RefXVC system with multi-reference technique while the source is excluded in the references during training.
    \item \textbf{NANSY}\cite{choi2022nansy}: A recent work that is similar work with ours. It uses content-dependent time-varying speaker embedding to improve the speaker identity. We trained the model with official implementations using the same dataset and Mel-spectrogram configuration.
    \item \textbf{Diff-HierVC}\cite{choi23d_interspeech}: A recent work employs SSL features to represent content information and utilizes the style encoder from \cite{pmlr-v139-min21b} to extract global speaker embeddings, with a diffusion model used as the generator.
\end{itemize}

 \begin{table*}
     \centering
     \caption{Results of MOS test on speech quality, CMOS test on speaker similarity and WER. English-to-Spanish denotes that the English source speech is converted into the voice of a Spanish speaker and vice-versa for Spanish-to-English. P-values are calculated between different systems and the baseline system.}
     \begin{tabular}{c|ccccc|ccccc} \toprule
        &  \multicolumn{5}{c}{English-to-Spanish} & \multicolumn{5}{|c}{Spanish-to-English} \\ \midrule
       System   &  quality & p-value & similarity & p-value & WER &  quality & p-value & similarity & p-value & WER \\ \midrule
        Baseline & 4.28\textpm0.07 & - & 3.86\textpm0.05 & - & 3.76& 4.3\textpm0.07 & - & 3.93\textpm0.05 & - & 4.37\\
        Diff-HierVC & 4.21\textpm0.03 & 0.007 & 3.91\textpm0.02 & \textless0.001 &  3.89& 4.23\textpm0.04 &  \textless0.001 & 3.92\textpm0.03 & 0.23& 4.52\\
        NANSY & 4.23\textpm0.06 & 0.002 & 3.98\textpm0.06 & \textless0.001 & 3.52& 4.33\textpm0.05 & 0.015& 4.11\textpm0.07 & \textless0.001 & 4.34\\
        Single-RefXVC  & 4.34\textpm0.08 &  \textless0.001 & 4.01\textpm0.05 & \textless0.001 & 3.18& 4.31\textpm0.07 & 0.14& 4.24\textpm0.07 & \textless0.001 & 4.21\\
        RefXVC (source-included) & 4.32\textpm0.06 & \textless0.001 & 4.24\textpm0.07 & \textless0.001 & 3.23& 4.35\textpm0.06 & \textless0.001 & 4.34\textpm0.05 & \textless0.001 & 4.17\\
        RefXVC (source-excluded) & \textbf{4.35}\textpm0.08 & \textless0.001 & \textbf{4.39}\textpm0.07 & \textless0.001 & \textbf{3.15}& \textbf{4.36}\textpm0.07 & \textless0.001 & \textbf{4.35}\textpm0.07 & \textless0.001 & \textbf{4.15}\\ 
        Ground-truth & 4.85\textpm0.05 & \textless0.001 & 4.94\textpm0.07 & \textless0.001 & 2.23& 4.91\textpm0.05 & \textless0.001 & 4.95\textpm0.06 & \textless0.001 & 3.58\\
        \bottomrule
     \end{tabular}
     \label{tab:mos}
 \end{table*}

We measured naturalness with a 5-scale mean opinion score (MOS [1-5]). Speaker similarity is also measured with a 5-scale comparison mean opinion score (CMOS [1-5]). We invited 20 native English speakers to conduct the listening experiments. We also calculate the Word Error Rate (WER) using Whisper-large \cite{pmlr-v202-radford23a} for both conversion pairs as an objective metric to evaluate the intelligibility. For all the experiments, the source speaker and target speaker were both unseen during training, which means we performed zero-shot any-to-any XVC. The results are presented in Table \ref{tab:mos}.

As shown in the table, our proposed systems outperformed the baseline, NANSY and Diff-HierVC in terms of both speech quality and speaker similarity. The PMN model improved the speaker similarity score over the baseline, and the multi-PMN systems further improved the scores by incorporating multiple references with or without the source speech. In addition, we have conducted comparisons using single speaker reference and multiple averaged speaker references with the baseline method. The performance differences are minimal, as evidenced by the visualized speaker embedding shown in Fig. 3(b), where the variance within the same speaker is minimal. The results demonstrate the effectiveness of our proposed system in XVC tasks.

\begin{figure}
    \centering
    \includegraphics[width=.45\textwidth]{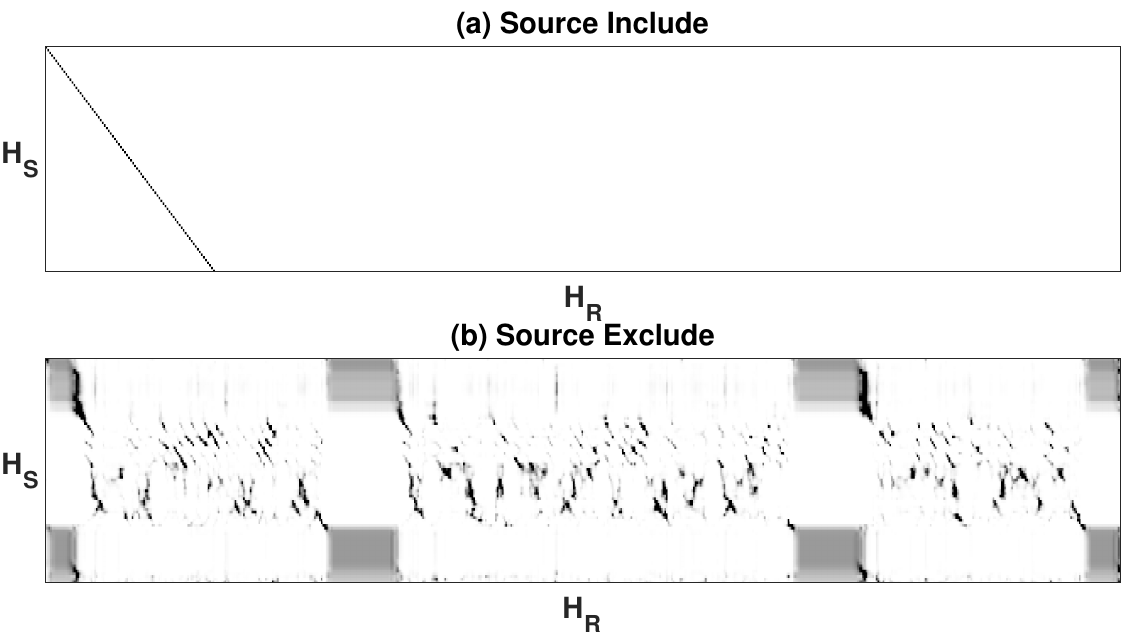}
    \caption{An illustration of the multi-reference alignment between Hs and Hr as computed by the PMN. (a) Source-included reference: The PMN focuses predominantly on the source utterance itself, as indicated by the concentration of attention weights along the diagonal. (b) Source-excluded reference: The PMN attends to the entire reference sentence, as evidenced by the more distributed pattern of attention weights across the heatmap.}
    \label{fig:att}
\end{figure}

We observed that the result obtained by excluding reference from the source is better than the one obtained by including it. Figure \ref{fig:att} illustrates an example of the alignment between $H_S$ and $H_R$ during training with different settings. In this example, we used three utterances as reference speech, with the first one being identical to the source in (a). This demonstrates that when the source is included during training, the pronunciation matching network tends to focus solely on the source utterance and disregards content information from other references. This contradicts the reason why we designed this mechanism. When excluding the source from the references, the pronunciation matching network can attend to the entire sentence and utilize the rich content information provided by the references. The figure confirms our conclusion that excluding the source is more effective in leveraging information from multiple references.

From the results presented in Table \ref{tab:mos}, there appear to be some differences in performance between converting Spanish to English and English to Spanish audio. While the differences are relatively small, they do indicate that there might be subtle challenges unique to each conversion direction. One potential difficulty in converting Spanish to English could stem from the distinct phonetic and prosodic characteristics of the two languages. Spanish, for example, has a more consistent syllable-timed rhythm and a relatively simpler vowel system compared to the more stress-timed rhythm and complex vowel system of English \cite{carter2005quantifying}. These differences can pose challenges for voice conversion models, which need to accurately capture and reproduce the nuanced phonetic and prosodic features of the target language.

 \begin{figure}
     \centering
     \includegraphics[width=.43\textwidth]{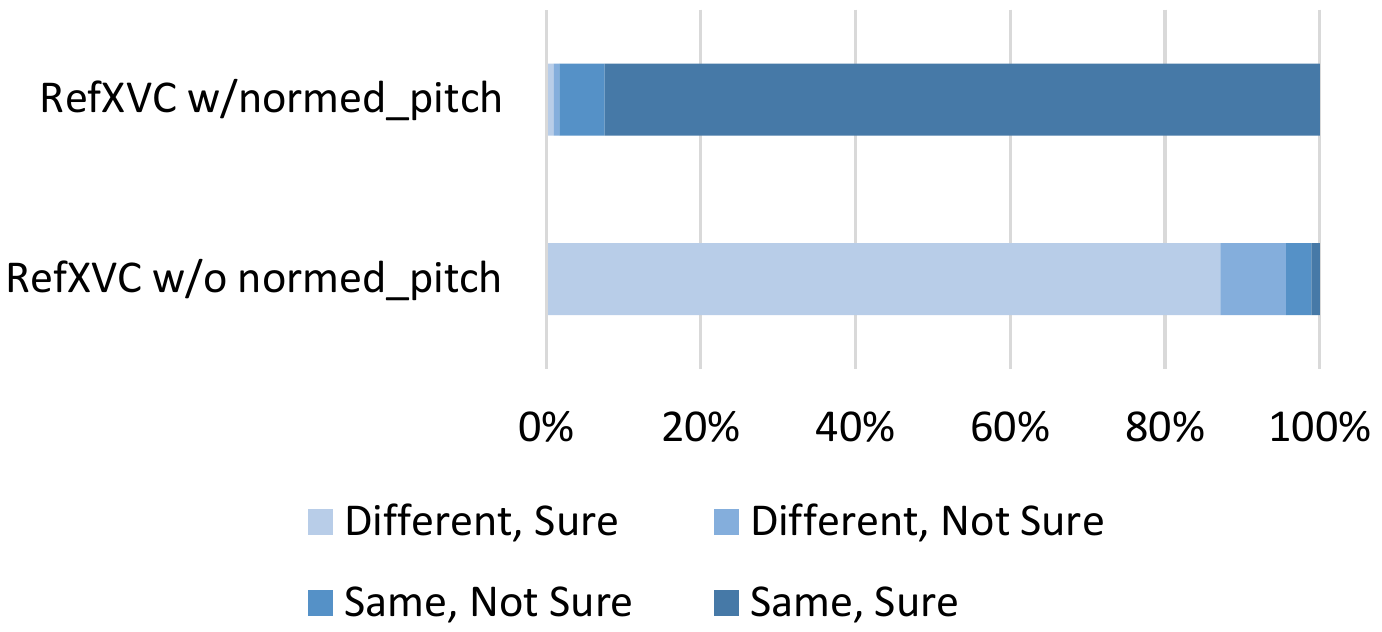}
     \caption{The prosody similarity test results between converted speech with and without using pitch normalization information. A higher percentage of `Same (not sure)' and `Same (sure)' together suggests a higher similarity to the natural source speech, which is more preferred.}
     \label{fig:pitch}
     \vspace{-3mm}
 \end{figure}

 \subsubsection{Evaluation on prosody}

In addition to the evaluation of speech quality and speaker similarity, we also conducted a subjective evaluation to assess the effectiveness of incorporating normalized pitch as an additional input to the RefXVC system. This evaluation aimed to investigate whether the inclusion of normalized pitch can improve the prosody of the converted speech, making it more similar to the source native speech.

We generated converted speech using both RefXVC systems, with and without the normalized pitch input, and compared the prosody between the converted speech and the source ground-truth speech. To assess the prosody similarity, we recruited 20 native English speakers to participate in the subjective listening tests. The participants were instructed to focus on only the prosody similarity, irrespective of speaker identity, and were provided with four response options: "Different, Sure", "Different, Not Sure", "Same, Not Sure", and "Same, Sure".

The results of the evaluation are presented in figure \ref{fig:pitch}. As shown in the figure, the RefXVC system with the normalized pitch input achieved a higher percentage of "Same, Sure" responses compared to the system without the normalized pitch input. This indicates that the normalized pitch input can effectively improve the prosody of the converted speech.

These results further demonstrate the importance of considering prosody in XVC and the effectiveness of leveraging reference information from the source to improve prosody. By incorporating normalized pitch as an additional input, RefXVC can improve the prosody of the converted speech, reducing foreign accent in XVC and making the speech more similar to the source native speech.


\section{Conclusion}
In this paper, we introduced a novel approach to cross-lingual voice conversion that maximizes reference leveraging in multiple ways. We proposed a timbre encoder and a pronunciation matching network to exploit the relationship between timbre and pronunciation in different languages and employed multiple reference sources to capture the tonal variations in a speaker's speech more accurately. Furthermore, we introduced the use of normalized pitch as an additional input to enhance the prosody of the converted speech and prevent foreign accents. Our experimental results demonstrate that our proposed approach outperformed state-of-the-art methods in terms of objective evaluation metrics, and subjective evaluation results confirmed the effectiveness of our approach in improving the naturalness and similarity of the converted speech. However, we acknowledge certain limitations in our work. The generalization of our method to unseen languages remains an area for further investigation. In future work, we plan to explore the generalization capabilities of RefXVC to a wider array of languages.


\bibliographystyle{IEEEtran}
\bibliography{sample-base}

\newpage




\vfill

\end{document}